\documentclass[11pt]{article}

\def\be{\begin{equation}}
\def\ee{\end{equation}}
\def\bea{\begin{eqnarray}}
\def\eea{\end{eqnarray}}
\def\bean{\begin{eqnarray*}}
\def\eean{\end{eqnarray*}}
\def\c{\cite}
\def\la{\label}
\def\r{\ref}

\begin{document}
\rightline{Jan 4 1999}
\vskip 1.5 true cm
\begin{center}
{\Large  The Camassa Holm Equation: \\[5pt]            
Conserved Quantities and the Initial Value Problem}
\vskip 1.5 true cm
{\large Michael Fisher$\;^1$\ \ and\ \ Jeremy Schiff$\;^2$}
\vskip 0.8 true cm
{\it $^1$ Department of Mathematics}\\
{\it University of Haifa, Mount Carmel, Haifa 31905, Israel}\\
{\small michael@eit.com}\\[5pt]
{\it $^2$ Department of Mathematics and Computer Science }\\
{\it Bar--Ilan University, Ramat Gan 52900, Israel}\\
{\small schiff@math.biu.ac.il}
\end{center}
\vskip 1.5 true cm
\begin{quote} 
\centerline{\bf Abstract}
Using a Miura-Gardner-Kruskal type construction, we show that the 
Camassa-Holm equation has an infinite number of {\em local} conserved
quantities. We explore the implications of these
conserved quantities for global well-posedness.
\end{quote}
\vskip 1 true cm

\noindent{\bf 1. Introduction.}

\smallskip

Much interest has been developing in the Camassa-Holm (CH) equation
\be m_t=-2mu_x-m_xu\ , \qquad m=u-u_{xx}\ . \la{CH}\ee
This equation first appeared in work of Fuchssteiner and Fokas \c{FF}
as an example of a bihamiltonian system, but more recently it was
rediscovered by Camassa and Holm \c{CH} as a model for shallow water 
waves. In addition, Misio{\l}ek \c{M} has shown that it describes a 
geodesic flow on the group Diff($S^1$). In these and other regards, 
the CH equation has much in common with the KdV equation, but there
are also a number of significant differences. In particular,
when considered as an evolution equation on a suitable Sobolev
space, KdV is {\em globally well-posed}, while CH is in general
not \c{CH,C,LO}; the first derivative of a solution of CH can become
infinite in finite time.

In this paper we present analogs for CH of two pieces of KdV theory.
The first concerns the construction of the conserved quantities.
Bihamiltonian structure implies that CH has an infinite 
number of conserved quantities. Explicitly, (\r{CH}) can be written 
either in the form
\bea
m_t &=& -B_1\frac{\delta H_2}{\delta m}\ , \\
B_1=\partial_x-\partial_x^3\ , && 
H_2=\frac12\int (u^3+uu_x^2) \ dx\ , \nonumber
\eea 
or in the form 
\bea
m_t &=& -B_2\frac{\delta H_1}{\delta m}\ , \\
B_2=\partial_x m + m \partial_x\ , && 
H_1=\frac12\int (u^2+u_x^2) \ dx\ . \nonumber
\eea 
Since $B_1$, $B_2$ are a {\em hamiltonian pair} \c{Olver},
the bi-infinite sequence of functionals  $\ldots,H_{-1},\linebreak[0]
H_0,\linebreak[0]H_1,\ldots$ defined by
\be B_2\frac{\delta H_n}{\delta m}=
    B_1\frac{\delta H_{n+1}}{\delta m}\ , \qquad n\in{\bf Z}\ , \la{CHlen}\ee
are  conserved quantities in involution with respect to 
the Poisson brackets determined by either $B_1$ or $B_2$. In \c{CH},
formulae  were given for $H_0,H_{-1},H_{-2}$, viz.
\be H_0=\int m\ dx\ , \qquad
     H_{-1}=\int \sqrt{m} dx\ , \qquad 
     H_{-2}=-\frac14\int\left(\frac{m_x^2}{4m^{5/2}}+\frac1{\sqrt{m}}
                  \right) \ dx\ \la{H-2}\ 
\ee
(there is a typo in the coefficients of $H_{-2}$ in \c{CH}). 
Unfortunately, however,
the need to invert either $B_1$ or $B_2$ each time (\r{CHlen}) is 
used makes it very hard to use this to generate further explicit formulae 
for the $H_n$, or to prove anything about them. 
The first result of this paper is an alternative
derivation of a bi-infinite sequence of conserved quantities for CH
(which we believe, but do not prove, to be equivalent to the $H_n$).
Our method shows directly that there are an infinite number of
conserved quantities that are {\em local}, i.e. integrals of some 
function of the fields $m$ and their $x$-derivatives. 
The existence of two constructions for the conserved quantities is 
familiar from KdV theory; for KdV the bihamiltonian structure 
gives the Lenard recursion \c{MGK5}, but locality is much easier to
show via the Miura-Green-Kruskal (MGK) construction \c{MGK2}.
The new derivation we give for a CH is precisely an analog of MGK
for KdV. 

Our second result concerns the relevance of the conserved quantities for 
the initial value problem. For KdV the relevant results are due
to Lax \c{Lax}, who showed, for the periodic problem, that the conserved
quantities bound Sobolev norms. We give a similar result for CH, 
restricted, however, to solutions with $m>0$. In this
case global well-posedness has been proved for CH \c{C}.

One more point needs to be explained in this introduction. In most
of the paper we do not work directly with the CH equation, but
rather with the {\em associated Camassa-Holm} (ACH) equation, introduced
in \c{me} and related to CH by a change of coordinates (see also \c{Hone}).
In section 2 we describe the relationship between CH and ACH, so that the 
results proved for ACH can be immediately translated into results for CH. 
In section 3, we give the MGK-type construction for ACH, proving there exist 
an infinite number of local conserved quantities of ACH and hence also of CH.
And in section 4, we show how these conserved quantities bound certain
norms for certain classes of solutions of ACH, and hence also for CH. 

\vfill\eject

\smallskip

\noindent{\bf 2. CH and ACH}

\smallskip

We concentrate our attention on two types of solution of CH,
(1) solutions with $m>0$ satisfying $m\rightarrow h^2$ as 
$\vert x\vert\rightarrow\infty$, where $h$ is a positive constant, 
and (2) solutions with $m>0$, and $u,m$ periodic in $x$, with
period independent of $t$. In general when $m>0$ we can define
$p=\sqrt{m}$, and the first equation of (\r{CH}) becomes $p_t=
-(up)_x$. This implies we can define new coordinates $t_0,t_1$ via
\be
dt_0 = p\ dx -pu\ dt\ , \qquad
dt_1 = dt \ .
\la{chcoor}\ee
More precisely, these define $t_0,t_1$ up to translations; choosing
the origin to coincide with the origin of $x,t$ coordinates we
have
\be 
t_0 = \int_0^x p(x',t)\  dx' - \int_0^t u(0,t')p(0,t')\ dt'\ ,\qquad
t_1 = t\ .
\ee
Transforming to the new coordinates gives the associated
Camassa-Holm (ACH) equation:
\be \dot{p}=-p^2u', \qquad u=p^2-p\left(\frac{\dot{p}}{p}
              \right)'\ .   \la{ACH} \ee
Here a prime denotes differentiation with respect to $t_0$, and
a dot differentiation with respect to $t_1$.
The two classes of solutions of CH introduced above correspond,
respectively, to (1) solutions of ACH with $p>0$, satisfying 
$p\rightarrow h$ as $\vert t_0 \vert \rightarrow \infty$, and
(2) solutions of ACH with $p>0$ and $p,u$ periodic in $t_0$,
with period independent of $t_1$. In the latter case, a solution of
CH with period $T$ corresponds to a solution of ACH with 
period $S=\int_0^T p(x,t)\ dx$ (which is independent of $t$), and
in the opposite direction a solution of ACH with period $S$ 
corresponds to a solution of CH with period $T=\int_0^S (1/p(t_0,t_1))
\ dt_0$ (which is independent of $t_1$). 

Evidently, a solution of CH in one of the two classes
under consideration exists for all $t$ if and only if the 
corresponding solution of ACH does. There is also a correspondence
between conserved quantities of the two equations (we thank Andy
Hone for explaining this to us, citing it as a  result
of Rogers, see \cite{Rog}). Suppose ${\cal X},{\cal T}$ are functions 
of $m,u$ and
their $x,t$ derivatives, such that ${\cal X}_t={\cal T}_x$ follows 
from (\r{CH}). 
Then $\int {\cal X}\ dx$ is a conserved quantity of CH. Using the relations 
$\partial_x=p\partial_{t_0}$ and 
$\partial_t=\partial_{t_1}-pu\partial_{t_0}$, ${\cal X}$ and 
${\cal T}$ can be
rewritten as functions of $p,u$ and their $t_0,t_1$ derivatives,
and it is straightforward to check that $\partial_{t_1}({\cal X}/p)=
\partial_{t_0}({\cal T}+u{\cal X})$. 
Thus $\int ({\cal X}/p)\ dt_0$ is a conserved
quantity of ACH. This procedure can be reversed, and we obtain a 
correspondence between conserved quantities of the two equations.
We will call a conserved quantity of CH (ACH) {\em local} if
it is of the form $\int {\cal X}\ dx$ ($\int \tilde{{\cal X}}\ dt_0$), 
where 
${\cal X}$ ($\tilde{{\cal X}}$) is a function of $m$ ($p$) and its 
$x$ ($t_0$)
derivatives alone. The general correspondence just described
can be checked to reduce to a correspondence of local conserved 
quantities.

The results we have just presented allow us to study ACH instead
of CH. From the form (\r{ACH}) of ACH  it is not
clear in what sense this is an evolution equation, but it turns
out that when we restrict to either of the classes of solutions
introduced above it can be written in a simple evolutionary form.
To see this we first eliminate $u$ from (\r{ACH}). Gathering all
the terms with $t_1$ derivatives on one side, the resulting equation 
can be written
\be
\left(\partial_{t_0}^2-\frac{p''}{2p}+\frac{p'^2}{4p^2}-\frac1{p^2}
\right) \frac{\dot{p}}{\sqrt{p}} = 2\sqrt{p}\ p'\ .
\ee 
To write this in evolutionary form (i.e. to have an explicit expression
for $\dot{p}$) we need to solve the second order 
ordinary differential equation 
\be 
\left(\partial_{t_0}^2-\frac{p''}{2p}+\frac{p'^2}{4p^2}-\frac1{p^2}
\right) y = f 
\la{2ode}\ee
(here $f,p$ are given and $y$ is the unknown).
For class (1) of solutions,  $f=2\sqrt{p}\ p'\rightarrow 0$  as 
$\vert t_0\vert\rightarrow\infty$, and we need a (hopefully unique) 
solution $y$ obeying a similar condition. Similarly,
for class (2) of solutions, $f$  is periodic, and we 
need a (hopefully unique) periodic solution $y$. Remarkably, for
arbitrary $p$ it is possible to explicitly solve the homogeneous
problem corresponding to (\r{2ode}) (i.e. the case $f=0$), and 
thence by standard methods solve the two inhomogeneous problems.
Explicitly, in the homogeneous case, (\r{2ode}) has two linearly 
independent solutions
\be y_{\pm}(t_0) = \sqrt{p(t_0)}
       \exp\left(\pm\int^{t_0} \frac{ds_0}{p(s_0)}\right)\ ,
\ee
and the two evolutionary forms of ACH are given by
\bea 
\dot{p}(t_0,t_1) &=& 
  -\int_{-\infty}^{\infty}p(t_0,t_1)p(s_0,t_1)p'(s_0,t_1)  
   \exp\left(-\left\vert\int_{s_0}^{t_0}\frac{du_0}{p(u_0,t_1)}
     \right\vert\right)
\ ds_0\  \la{infCH},\\
\dot{p}(t_0,t_1) &=& 
  -\int_0^Sp(t_0,t_1)p(s_0,t_1)p'(s_0,t_1)
\frac
{\cosh 
\left(
\frac12 \int_{0}^{S}\frac{du_0}{p(u_0,t_1)}
-\left\vert\int_{s_0}^{t_0-\left[\frac{t_0}{S}\right]S 
      }\frac{du_0}{p(u_0,t_1)}\right\vert
\right)
}
{\sinh
\left(
\frac12 \int_{0}^{S}\frac{du_0}{p(u_0,t_1)}
\right)
}\ ds_0
\la{perCH}\eea
respectively. In the latter, $S$ denotes the period of $p$ as a
function of $t_0$, and $\left[\frac{t_0}{S}\right]$ denotes
the largest integer not exceeding $\frac{t_0}{S}$; it is 
an enjoyable exercise to check that the expression $\cosh(\ldots)$
appearing in (\r{perCH}) is continuous when $t_0$ is an integer 
multiple of $S$, for any $s_0$.

\smallskip

\noindent{\bf 3. The MGK construction for ACH}

\smallskip

It was shown in \c{me} that (\r{ACH}) has a strong B\"acklund transformation 
\be 
p\rightarrow p-2s', \qquad u\rightarrow u+\frac{2\dot{s}}{p(p-2s')}\ ,
\ee
where $s$ satisfies 
\bea
s' &=& -\frac{s^2}{2p\lambda} + \frac{\lambda}{2p}+\frac{p}2  \ , 
   \la{BT1}\\
\dot{s} &=& -s^2 +\frac{s\dot{p}}{p} + \lambda(\lambda+u) \  \la{BT2}
\eea
($\lambda$ is a parameter). It is straightforward to check that 
\be \dot{\left({\frac{s}{p}}\right)}=
\lambda\left(2s -{\frac{\dot{p}}{p}}\right)'\  ,
\ee
from which we deduce that $\int (s/p) \ dt_0$ is a conserved quantity. But
$s$ is dependent on $\lambda$; so if we can find a consistent expansion of
$s$ in powers of $\lambda$, then in fact each term in the expansion of
$\int(s/p)\  dt_0$ will yield a conserved quantity. Two such
consistent expansions of $s$ are:
\be
s = \sum_{n=1}^{\infty} s_n\lambda^{\frac{n}2}~ \qquad {\rm with~} s_1=p
\la{firstser}\ee
and
\be
s = \sum_{n=0}^{\infty} r_n\lambda^{1-n} \qquad {\rm with~} r_0=1 \ .
\la{secser}\ee
Using (\r{firstser}) and  substituting in (\r{BT1}) gives the recursion
\be
s_{n+1}=-s_n'+\frac1{2p}\left(\delta_{n2}-\sum_{i=0}^{n-2}
s_{i+2}s_{n-i}
\right)\qquad n=1,2,\ldots ~~.
\la{srec}\ee
With the aid of a symbolic manipulator it is easy to compute the first 
few of the $s_n$; we find that $s_n/p$ is a total derivative for $n$
even, but that for $n$ odd we obtain nontrivial conserved quantities 
(both these statements can easily be proved). Writing 
$K_n=\int(s_{2n-1}/p)\ dt_0$, $n=2,3,\ldots$, and integrating by parts to 
reduce the order of derivatives appearing, we find (up to unimportant 
overall constants):
\bea
K_2 &=& \int \left(\frac{p'^2}{p^2} + \frac{1}{p^2}\right)\ dt_0\ ,
            \nonumber\\
K_3 &=& \int \left(4\frac{p''^2}{p^2} - 3\frac{p'^4}{p^4} +10\frac{p'^2}{p^4}
         + \frac{1}{p^4}\right) \ dt_0\ ,\\
K_4 &=& \int \left(
    24\frac{p'''^2}{p^2}+72\frac{p''^3}{p^3}-228\frac{p'^2p''^2}{p^4}
  +135\frac{p'^6}{p^6}+84\frac{p''^2}{p^4}-259\frac{p'^4}{p^6}
  +105\frac{p'^2}{p^6}+\frac{3}{p^6}
    \right)\ dt_0\ .\nonumber
\eea
Following the procedure of converting conserved quantities of
ACH to those of CH, as described in section 2 , it can be seen that 
$K_2$ gives $H_{-2}$ as given in (\r{H-2}) up to an overall
constant. Similarly, lengthy calculations show that the conserved
quantity of CH determined by $K_3$ is actually $H_{-3}$, 
up to an overall constant; the explicit form of $H_{-3}$ is
\be
H_{-3}=\frac18\int\left( \frac{m_{xx}^2}{m^{7/2}}  
                      -  \frac{35}{16}\frac{m_x^4}{m^{11/2}}
                      +  \frac52\frac{m_x^2}{m^{7/2}}
                      +\frac1{m^{3/2}}
                  \right) \ dx\  .\la{H-3}\ 
\ee
We conjecture that in fact $K_n$ is equivalent to $H_{-n}$ for 
all $n=2,3,\ldots$ (up to multiplication by overall constants). 
In any case, it is clear that the $K_n$ are local, and therefore so are 
the corresponding conserved quantities of CH. 

Using now the series (\r{secser}) for $s$ and substituting in (\r{BT2})
gives the recursion
\be
r_{n+1}=\frac12\left(u\delta_{n0}-\dot{r_n}+\frac{r_n\dot{p}}{p}-
   \sum_{i=0}^{n-1} r_{i+1}r_{n-i}\right) \qquad n=0,1,\ldots ~~.
\ee
Explicit expressions for the first few $r_n$ are:
\bea
r_1 &=& \frac12\left(u+\frac{\dot{p}}{p}\right)\ , \nonumber\\
r_2 &=& \frac18\left( \left(\frac{\dot{p}}{p}\right)^2 - 2\dot{u}
            -2 {\left(\frac{\dot{p}}{p}\right)}^{\cdot} - u^2 \right)\ ,  \\
r_3 &=& \frac1{16}\left( 2{\left(\frac{\dot{p}}{p}\right)}^{\cdot\cdot}
    + 2\left(u-\left(\frac{\dot{p}}{p}\right)\right)
     {\left(\frac{\dot{p}}{p}\right)}^{\cdot}
    -u\left(\frac{\dot{p}}{p}\right)^2+2\ddot{u}+4u\dot{u}+u^3
   \right)\ .\nonumber
\eea
The resulting conserved quantities $\int (r_n/p)\ dt_0$ are in general 
not local, since these expressions involve $t_1$-derivatives. 
$r_0$ gives rise to the conserved quantity $T=\int (1/p)\ dt_0$ of ACH 
mentioned in section 2. Lengthy 
calculations, not reproduced here, show that the conserved quantities
of CH determined by $r_1,r_2,r_3$ are $H_0,H_1,H_2$ respectively
(up to multiplication by overall constants). We conjecture that
$r_n$ gives rise to the conserved quantity $H_{n-1}$ of CH for all $n>0$.

In the next section we will use the first series of conserved quantities,
and it will be useful to have a number of facts about them available.
The following results are easy to prove:
\begin{enumerate}
\item For $n>1$, $s_n$ is a polynomial in $\frac1{p},p',p'',\ldots$; 
   $s_n$ is odd under $p\rightarrow -p$.
\item Assigning weight $1$ to $\frac1{p}$ and weight $n-1$ to $p^{(n)}$,
   $s_n$ is a sum of terms of weight $n-2$.
\item Each term in $s_n$ contains at most $n-1$ derivatives (so, for
  example, in $s_7$ a $(p''')^2/p$ term is allowed, but not a
  $(p''')^2(p')^2/p$ term).
\end{enumerate}
These results concern $s_n$. In constructing the conserved quantities
$K_n=\int (s_{2n-1}/p)\ dt_0$ ($n\ge 2$) we are allowed to integrate by 
parts to reduce the order of derivatives appearing. In general, we can 
continue to do this until the highest derivative appearing in any term
appears nonlinearly. After this, we have  $K_n=\int {\cal K}_n\ dt_0$,
where:
\begin{enumerate}
\item ${\cal K}_n$ is a polynomial in $\frac1{p},p',p'',\ldots$,
    even under $p\rightarrow -p$ and 
    divisible by $\frac1{p^2}$.
\item Assigning weight $1$ to $\frac1{p}$ and weight $n-1$ to $p^{(n)}$,
   ${\cal K}_n$ is a sum of terms of weight $2n-2$.
\item Each term in ${\cal K}_n$ contains at most $2n-2$ derivatives, and
   in each term the highest derivative appears nonlinearly.
\end{enumerate}
It  follows that the highest order derivative appearing in ${\cal K}_n$
is $p^{(n-1)}$, and this appears only in a term proportional to
$(p^{(n-1)})^2/p^2$. We will assume in what follows that the coefficient
of this term is always nonzero, in which case we can without loss
of generality take it to be positive. ${\cal K}_n$ also has a term with 
no derivatives, proportional to $1/p^{2n-2}$. Since this term is not
affected in any way by the procedure of integration by parts, it also
appears in $s_{2n-1}/p$, and from the recursion (\r{srec}) it is
possible to show its coefficient is nonzero. This guarantees the 
nontriviality of the series of conserved quantities. But it also implies
that the $K_n$ as we have defined them are actually infinite in the 
case of solutions with $p\rightarrow h$ as $\vert t_0\vert \rightarrow
\infty$. This can be rectified by adding a suitable constant to 
${\cal K}_n$; for example $K_2$ should be modified to 
$ \int_{-\infty}^{\infty}
 \left(\frac{p'^2}{p^2} + \frac{1}{p^2} 
  - \frac1{h^2} \right)\ dt_0\ $.
Similarly the conserved quantity $T$ should be modified to
$\int_{-\infty}^{\infty} \left(\frac1{p} - \frac1{h}\right) \ dt_0$.

\smallskip

\noindent{\bf 4. Bounds on norms from the conserved quantities}

\smallskip

In this section we consider only the periodic case of ACH. By rescaling
$p,t_0,t_1$ we can without loss of generality take $S=1$. The $p>0$
condition is awkward to work with, so we eliminate it via
the substitution $p(t_0,t_1)=e^{v(t_0,t_1)}$. The evolution equation
(\r{perCH}) becomes 
\be
\dot{v}(t_0,t_1) =
  -\int_0^1 e^{2v(s_0,t_1)}v'(s_0,t_1)
\frac
{\cosh 
\left(
\frac12 \int_{0}^{1} e^{-v(u_0,t_1)}\ du_0
-\left\vert\int_{s_0}^{t_0-\left[t_0\right]} e^{-v(u_0,t_1)}\ du_0 
 \right\vert
\right)
}
{\sinh
\left(
\frac12 \int_{0}^{1} e^{-v(u_0,t_1)}\ du_0
\right)
}\ ds_0\ .
\la{vflow}\ee
Usual contraction mapping 
methods can be used to prove the local existence of solutions for
this equation. For convenience we write out the first few local 
conserved quantities in terms of the new field $v$:
\bea
T&=& \int_0^1 e^{-v} \ dt_0\ , \nonumber\\
K_2&=& \int_0^1 \left((v')^2 + e^{-2v}\right)\ dt_0\ , \nonumber\\
K_3&=& \int_0^1 \left(4(v'')^2 + (v')^4 + 10e^{-2v}(v')^2 + e^{-4v}
                      \right)\ dt_0\ , \\
K_4&=& \int_0^1 \left(24(v''')^2 + 60(v')^2(v'')^2+3(v')^6 
                    \right.\nonumber\\
&& ~~~~~~~~~\left.+ 
        e^{-2v}\left(84(v'')^2-63(v')^4\right)
       + 105 e^{-4v} (v')^2 + 3e^{-6v}\right)\ dt_0 \ .\nonumber
\eea
To obtain these formulas we have performed some integration by
parts, so that once again all terms in the densities have their
highest derivative appearing nonlinearly. A direct check that $T$ is 
conserved under the flow (\r{vflow}) is a long but ultimately rewarding 
exercise. The following analog of Lax's theorem now holds:

\smallskip

\noindent{\it Theorem:} Let $v$ be a smooth function of period $1$,
and $n$ a positive integer; the quantities 
$$ {\rm max}\{\vert v(t_0) \vert,
               \vert v'(t_0) \vert, \ldots,
               \vert v^{(n-1)}(t_0) \vert\}, \qquad 
\int (v^{(n)})^2\ dt_0
$$
can be bounded in terms of $T,K_2,\ldots,K_{n+1}$.

\smallskip

\noindent{\it Proof:} The cases $n=1,2$ needs to be checked 
individually; for larger $n$ we can proceed by induction, exactly
as in Lax \c{Lax}, because for $n\ge 4$ the density of $K_n$ is at 
most quadratic not only in $v^{(n-1)}$ but also in $v^{(n-2)}$.
(For $K_4$ this is evident from the explicit formula, but is 
fortuitous because there are no {\it a priori} reasons to exclude
a term proportional to $(v'')^3$; for $n>4$ it follows since all 
terms in the density of $K_n$ can have at most $2n-2$ derivatives.)

For $n=1$, writing $Q=\int_0^1 (v')^2 \ dt_0$,
we evidently have $Q< K_2$. Also, since for 
any $t_0,s_0$ 
$$ v(t_0)=v(s_0) + \int_{s_0}^{t_0} v'(u_0) \ du_0\ , $$
it follows from the Schwarz inequality that 
$$ (v(t_0))^2 \le 2 \left( (v(s_0))^2 + Q \right) \ . $$
Taking $s_0$ to be any point such that $e^{-v(s_0)} = \int_0^1 e^{-v(u_0)}
\ du_0 = T$ (such a point exists by the mean value theorem), we have
at once that for all $t_0$
$$ (v(t_0))^2 \le 2 \left((\ln T)^2 + Q  \right)
              \le 2 \left((\ln T)^2 + K_2 \right)\ . $$
This gives a bound for ${\rm max}\{\vert v(t_0) \vert\}$. Moving 
now to $n=2$, again it is clear that $Q_2\equiv\int_0^1 (v'')^2 \ dt_0
\le K_3/4$, and an argument similar to that given before shows that for
all $t_0,s_0$ 
$$ (v'(t_0))^2 \le 2 \left( (v'(s_0))^2 + {\textstyle{
\frac{K_3}{4}}} \right) \ . $$
Choosing $s_0$ such that $(v'(s_0))^2=Q<K_2$ gives a bound on
${\rm max}\{\vert v'(t_0) \vert\}$. 
$\bullet$

\smallskip

Thus in particular we see how the conserved quantities of (\r{vflow})
give bounds on the Sobolev norms of solutions; in this sense the 
integrability of CH can be seen to be directly related to global 
well-posedness in the case $m>0$.

\end{document}